\documentclass[]{aastex701} 

\usepackage[utf8]{inputenc}  
\usepackage{amsmath}         
\usepackage{amsfonts}        
\usepackage{amssymb}         
\usepackage{cleveref}
\usepackage{bm}
\usepackage[inkscapelatex=false]{svg}
\usepackage{diagbox}

\usepackage{physics}         

\crefname{figure}{figure}{figures}

\begin{document}


\title{Ejected Surface Regolith as a Potential Source Material for Centaur Rings}

\correspondingauthor{Kaustub P. Anand}
\email{anand43@purdue.edu}

\author[0009-0007-2467-0139]{Kaustub P. Anand}
\affiliation{Department of Physics and Astronomy, \\
Purdue University, \\
525 Northwestern Ave, West Lafayette, IN 47907, USA}
\affiliation{Department of Earth, Atmospheric, and Planetary Science, \\
Purdue University, \\
550 Stadium Mall Drive, West Lafayette, IN 47907, USA}
\email{anand43@purdue.edu}

\author[0000-0003-1656-9704]{David A. Minton}
\affiliation{Department of Earth, Atmospheric, and Planetary Science, \\
Purdue University, \\
550 Stadium Mall Drive, West Lafayette, IN 47907, USA}
\affiliation{Department of Physics and Astronomy, \\
Purdue University, \\
525 Northwestern Ave, West Lafayette, IN 47907, USA}
\email{daminton@purdue.edu}

\author[0000-0003-2627-335X]{Julie Brisset}
\affiliation{Department of Physics, University of Central Florida, 4111 Libra Dr, Orlando FL 32816, USA}
\affiliation{Florida Space Institute, University of Central Florida, 12354 Research Pkwy, Orlando
FL 32826, USA}
\email{julie.brisset@ucf.edu}

\begin{abstract}
Ring systems have been observed around Centaur Chariklo~(10199) and other small bodies but their origin and dynamical histories are still debated. These small body ring systems challenge conventional models for the origin of planetary rings, especially when considering Centaurs’ often erratic cometary activity, their non-spherical shapes, and their relatively short dynamical lifetimes ($\sim 10^7$ years). A collisional origin for these rings is disfavored based on the low probability of collisions within their lifetimes, and so their mechanism of formation remains an open question. In this work, we use Swiftest, a N-body integrator with collisional fragmentation and higher-order gravitational harmonics, to test a hypothesis that rings could be formed from regolith ejected from a cometary outburst that is subsequently captured into a stable orbit. We show that ejected surface regolith is captured in orbit around ellipsoidal Centaurs like Chariklo and Chiron to form a proto-ring disk for at least $100$~rotations. This captured disk may serve as a starting point that can evolve into observed ring systems. Inter-particle collisions and the ellipsoidal gravity field facilitate this capture. Among the tested scenarios, a landslide or avalanche-like ejection from the equatorial plane shows the highest rate of capture, $\sim 30 - 90\%$ depending on the initial ejection parameters. This implies that rings could be an indicator of past activity on a Centaur and may be a more common feature among Centaurs depending on their shape and frequency of outbursts.
\end{abstract}


\section{Introduction} \label{sec:intro}
Ring systems have been found around Centaur Chariklo (10199)~\citep{Braga-ribas-2014}, Trans-Neptunian Objects (TNOs) Haumea (136108)~\citep{Ortiz2017TheOccultation} and Quaoar (50000)~\citep{Pereira2023TheQuaoar, Morgado2023Quaoar} via stellar occultation surveys. The rings' origin is unknown and they are the only confirmed solar system ring systems that are not around the giant planets. Dense material features have also been observed in the coma of Centaur 95P/(2060) Chiron but their status as conventional rings is debated because of their tentative and transient behavior, with spherical shells of material as an alternative hypothesis~\citep[][etc.]{Ortiz2023The15, Ortiz2015PossibleChiron,Sickafoose2023MaterialOccultation,Sickafoose2020Characterization2011,Pereira2025TheSystem,Ruprecht201529Features}. The hypotheses for ring formation vary from tidal disruptions during close planetary encounters~\citep{Hyodo2016FORMATIONENCOUNTERS} to debris being ejected in impact events~\citep{Charnoz2018TheSystems} to three-body interactions between the primary body and two smaller bodies~\citep{Melita2017}. Of the TNOs that are presently known to host rings, only Haumea shows evidence of a collisional history~\citep{Brown2007ABelt}. \citet{Melita2017} concluded that none of the proposed extrinsic mechanisms for forming rings are likely for the much smaller Centaur (10199) Chariklo. This suggests that some kind of intrinsic process may be responsible for generating their ring.

Centaurs are known to show periodic cometary activity \citep[][and others]{Bauer2008, Jewitt2024TheContinuum, Kareta2019Physical174P/Echeclus, Kareta2025Activity-induced20172022, Meech1990TheChiron, Lisse202229P/SchwassmannWachmannComets} and have short rotation periods~\citep[e.g.: $T_{Chariklo} = 7.004$~h,][see \cref{tab:centaurs}]{Fornasier2014TheActivity}. Due to their rapid rotations, they are also expected to have ellipsoidal shapes~\citep{Leiva2017, BragaRibas2023Chiron}. In this paper, we show that surface regolith ejected from a rapidly-rotating Centaur can be trapped in orbit in sufficient quantity to be potential source material for the observed ring systems. Regolith ejected by activity (such as an outburst or a landslide) has a large velocity component due to the Centaur's rotation. As the regolith particles leave the rapidly rotating non-inertial frame of the Centaur, the centrifugal contribution is $\sim 0.45$~v$_{escape}$ from the longest axis in equatorial plane for Chariklo, where v$_{escape}$ is the escape velocity of the Centaur (here Chariklo (10199)). Regolith particles then undergo collisions with other regolith particles within the gravitational field of the ellipsoidal Centaur, which act to stabilize the particles into bound orbits. In this work, we model the combined effects of higher-order gravitational harmonics and inter-particle collisional fragmentation on populations of particles representing material ejected from the surface of a Centaur. We show that particles that are initially on either suborbital or escaping trajectories can be altered into at least quasi-stable orbits. Such orbits could eventually lead to rings as interparticle collisions remove orbital energy, although in this work we do not follow the evolution of the captured particles' orbits all the way to true planar rings.

In this work, we used the integrators Regular Mixed Variable Symplectic~\citep[RMVS,][]{Levison1994TheComets} and Symplectic Massive Body Algorithm~\citep[SyMBA,][]{Duncan1998AENCOUNTERS}, as well as the collisional fragmentation model Fraggle, that are included in Swiftest \citep{Wishard2023Swiftest:Systems}. We have also added a gravitational harmonics routine for computing the non-spherical potential of Centaurs using the Spherical Harmonic TOOLS (SHTOOLS) library~\citep{Wieczorek2018SHTools:Harmonics}. This paper is organized as follows. In section \ref{sec:background}, we discuss relevant background information. In section \ref{sec:methods}, we describe the methods and simulation set up. In section \ref{sec:results} we show results of the simulations, and discuss the implications of the results in section~\ref{sec:discussion} with conclusions in section~\ref{sec:conclusions}. 

The vector notation in this paper may be different from what is typical in the field. A bold quantity such as $\mathbf{x}_{vec}$ indicates a vector, $x_{vec}$ indicates only the magnitude of the vector, and $\mathbf{\hat{x}}_{vec}$ indicates only the direction component of the vector. For our purposes, ``capture" is defined as the amount of regolith in bound orbit around the Centaur with eccentricity $e < 1$ and periapsis $q > 1$~R$_{Centaur}$ at a given moment of time. R$_{Centaur}$ is the volume equivalent radius of the Centaur in our simulations. We replace the subscript with the name of the Centaur when referring to a specific body. For example, R$_{Chariklo}$ is the volume equivalent radius for Chariklo (10199).

\section{Background} \label{sec:background}

\subsection{Observed Ring Properties} \label{sec:current rings}

(10199) Chariklo was the first small body to be discovered with rings \citep{Braga-ribas-2014}. Since then, rings have been discovered around TNOs (136108) Haumea~\citep{Sicardy2019RingHaumea} and (50000) Quaoar~\citep{Pereira2023TheQuaoar, Morgado2023Quaoar}, and dense material in the coma around the Centaur 95P/(2060) Chiron~\citep{Sickafoose2020Characterization2011,Sickafoose2023MaterialOccultation,Ortiz2023The15,Ortiz2015PossibleChiron}.
In this work, we will focus on the systems around the two Centaurs Chariklo and Chiron. All the above systems were observed via stellar occultation surveys with the observed ring parameters and locations for Chariklo in table~\ref{tab:ring parameters}. 

Focusing on the Centaurs, Chariklo's ring system contains two rings with a narrow gap between them \citep{Braga-ribas-2014}. The gap between the rings is expected to be from shepherding satellites, but these satellites have not been observed yet because they are expected to be about a few km in size and would be very difficult to observe~\citep{Braga-ribas-2014, Sickafoose2024NumericalPerturber}. However, an eccentricity gradient within the rings themselves could potentially maintain this gap~\citep{Hahn2023NbodyRinglets, Murray2000SolarDynamics}. Lastly, the ring system is assumed to be about the equatorial plane of Chariklo and has a mass corresponding to a $\sim1$~km-sized icy body~\citep{Braga-ribas-2014}. Using a surface density of $30 - 100$~g/cm$^\mathrm{2}$~\citep{Braga-ribas-2014} and ring properties from \citet{Morgado2021Refined2020}, we can estimate the total C1R ring mass to be $\sim 5 - 16 \times 10^{13}$~kg. We use a starting ring mass of $\sim 8 \times 10^{13}$~kg in our simulations.

Chiron's system is different from the ones above because its status as traditional rings is debated. Chiron is a known active Centaur with regular cometary activity~\citep{Meech1990TheChiron, Luu1990CometaryChiron, Jewitt2009TheCentaurs, Dobson2024TheChiron, Belskaya2010PolarimetryChariklo, Bus20012060Aphelion}. 
Unlike the other small bodies, it has dense material, sometimes interpreted as 2 or 3 ring-like structures or spherical shells~\citep{Ortiz2015PossibleChiron, Ortiz2023The15, Pereira2025TheSystem}, embedded in a debris disk/coma~\citep{Sickafoose2020Characterization2011, Sickafoose2023MaterialOccultation}. However, these dense material features have changed over time with varying widths, optical depths, and distances across multiple observational epochs~\citep{Sickafoose2020Characterization2011, Sickafoose2023MaterialOccultation, Ruprecht201529Features}. Thus, Chiron is an extremely complicated system of an active Centaur with dense material around it and no clear explanation for these features. With this in mind,
we also test Chiron in our simulations for two reasons: it is typically included in discussions about small body ring systems and it is a well-studied body that is applicable to our aim of testing orbit capture of ejected regolith particles.

The rings of Chariklo pose important questions regarding their origins and long-term dynamic stabilities. Centaurs' small sizes, short lifetimes of 1--10~Myr~\citep{Tiscareno2003TheCentaurs}, and lack of known collisional histories challenge conventional ring formation scenarios which typically involve tidal breakup or capture~\citep{Melita2017}. This is further discussed in section~\ref{sec:ring source hypotheses}. The ring systems' stabilities and longer term dynamics are also not well understood. Centaurs experience multiple close encounters with the giant planets over their lifetimes. These close encounters have a low probability to disrupt or largely alter their rings for a Chariklo-like orbit~\citep{Araujo2016THEPLANETS, Wood2017TheRings}. While the probability is also low for Chiron, it is about 6 times higher than that for Chariklo~\citep{Araujo2018RingsChiron}. 

The rings of Chariklo \citep{Leiva2017, Morgado2021Refined2020}, Haumea~\citep{Ortiz2017TheOccultation}, and Quaoar~\citep{Morgado2023Quaoar, Pereira2023TheQuaoar} seem to lie close to the 3/1 spin-orbit resonance (3/1 SOR)~\citep{Sicardy2021RingsKey}. The rings of Chariklo, Quaoar, and the outer feature of Chiron lie outside the Roche limit~\citep{Sicardy2025OriginsSystem, Pereira2023TheQuaoar, Morgado2023Quaoar}. However, this limit depends on a lot of factors such as mass distribution and density of a central body, and can be explained better by Roche density~\citep{Tiscareno2013CompositionsDensity}. The Roche density at a given distance is a threshold density. At a location, material density above this threshold is expected to coalesce into and be dominated by a single body such as a moon. Material density below the threshold will be broken up and is expected to stay dispersed like a ring. A density of $0.85$~g cm$^{-3}$ for Chariklo~\citep{Morgado2021Refined2020} gives us a Roche density of $\sim 0.16$~g cm$^{-3}$ at the location of the rings today. This is extremely low, even for porous icy material, and we would expect the ring material to clump together. While there may be some internal clumping, it is not observed yet for the aforementioned rings. It is not well understood how these rings are able to exist outside the Roche limit or the significance of the 3/1 SOR yet. Work is being done in the field to understand the rings' longer-term dynamics, but this is not within the scope of this project.

\subsection{Hypothesized Ring Formation Scenarios} \label{sec:ring source hypotheses}

In this section, we summarize the analysis of \citet{Melita2017}. They evaluate known ring formation scenarios for Chariklo, and demonstrate why they are unlikely sources for the rings we see today. Proposed ring formation mechanisms invoke tidal breakup, capture of a secondary smaller object, an impact on the Centaur, or a combination of them. They use values of $A = B = 117$~km and $C = 122$~km for the semi-axes of Chariklo. 

For a tidal breakup of an existing satellite, they conclude that the potential satellite densities, sizes, and timescales do not satisfy observational constraints. The Roche limit distance varies with the satellite's sphericity ($C / A$) and density. At higher sphericity values $\gtrsim 1.6$, the Roche limit asymptotes. From \citet{Melita2017}'s figure 1, the satellite has to take extreme values of density and sphericity for the Roche limit to exist at the ring location. A satellite of a similar density to Chariklo would have to have a radius of $4 - 7$~km and a low friction angle to break up at $\sim 3.6$~R$_{{Chariklo}}$ close to the current rings' location ($\sim 3.4$~R$_{Chariklo}$ in \citet{Melita2017} when using $R_{Chariklo} = 118$~km). If the satellite is tidally locked, the breakup distance is independent of friction angle~\citep{Holsapple2008TidalSystem} and needs to have a density of about $1/40$~$\rho_{{Chariklo}}$ for breakup at the ring location. Here, $\rho_{{Chariklo}}$ is the density of Chariklo in their work. If the density of the satellite is similar to that of Chariklo, then the breakup distance is at about $1$~R$_{{Chariklo}}$~\citep{Melita2017}.

Next, \citet{Melita2017} look at a satellite's orbital decay due to tides. The ring location ($\sim 400$~km) is inside the synchronous orbit ($\sim 696$~km). For tides raised by the Centaur, the timescale of inward evolution for a satellite would be $10^{5}$ -- $10^{14}$~yr for satellite radius $R = 100 $--$1$~km respectively, depending on starting location.
For tides raised on the Centaur, a satellite at the ring location would be pushed away from Chariklo at this location. Assuming Chariklo spun more slowly at an earlier time, the timescales to reach the current location would be $10^{5}$ -- $10^{11}$~yr for satellite radius $R = 100$ -- $1$~km respectively, depending on starting location. The timescales of orbit transfer for a $1$~km-sized icy body exceed that of Centaur lifetimes (on the order of $10^{6}$ -- $10^{7}$~yr), rendering a tidal transport and breakup origin improbable.

As a tidal origin is ruled out, \citet{Melita2017} simulate a 3-body encounter where two field objects collide in a perfectly inelastic collision, lose enough energy in the collision to become trapped in orbit around Chariklo, and then eventually break up at the ring location. For this to occur, they conclude that the bodies have radii of $6.5$~km and $330$~m with a relative impact velocity of $3$~km s$^{-1}$. The combined body would have a size similar to the bigger one. From the tidal calculations, the larger body itself would not breakup in time, and thus a 3-body encounter is also implausible.

A cratering event on Chariklo could eject debris, but is also unlikely to form the rings. About $10^{-2}$ -- $10^{-3}$ of the ejected mass would be bound in orbit. For an impact on Chariklo that gives us the observed ring mass, the impactor radius would need to be $0.2 - 1$~km and the expected crater would be $20 - 50$~km in diameter \citep{Melita2017}. At a higher resolution, this crater would be visible. However, using a scaling from \citet{Levison2000PlanetaryComets} the impact rate would be about $2.2 \cross 10^{-8}$~yr$^{-1}$, making this event improbable. If the impact occurred in the TNO region, the impact rate would be smaller at about $1.3 \cross 10^{-12}$~yr$^{-1}$ \citep{Melita2017}.

Lastly, a catastrophic impact on an existing satellite could also form debris for a ring. The impact would have to be on at least a km-sized body. This satellite would have a cross section of about $10^{-4}$ that of Chariklo. This gives us impact rates of about $10^{-12}$~yr$^{-1}$ in the Centaur region and $10^{-14}$~yr$^{-1}$ in the TNO region. Therefore, this event is also highly unlikely \citep{Melita2017}.

\begin{deluxetable}{|c|c|cc|c|}
\tablecaption{Ring parameters for Centaur Chariklo (10199).\label{tab:ring parameters}}
\tablehead{
\colhead{Centaur}&\colhead{Ring Width (km)}&\colhead{Ring Radius (km)} & \colhead{(R$_\mathrm{Centaur}$)}&\colhead{Source}}
\startdata
         (10199) Chariklo &$7$&  $391$ & $3.17$& \citet{Braga-ribas-2014}\\ 
         \cline{2-4} 
                         &$3$& $405$ & $3.28$&\\ 
\enddata
\tablecomments{Here we use $R_{Chariklo}=123.35$~km.}
\end{deluxetable}

\begin{deluxetable*}{|c|c|c|c|c|}
    \tablecaption{Physical characteristics of noted Centaurs. \label{tab:centaurs}}
    \tablehead{\colhead{Centaur} & \colhead{Mass (kg)} & \colhead{Semi-Axes ($A \times B \times C$) (km)} & \colhead{Escape Velocity (m/s)} & \colhead{Rotation Period (h)}}
    \startdata
    (10199) Chariklo&  $6.1 \cross 10^{18}$~$^\mathrm{L}$&  $157 \cross 139 \cross 86$~$^\mathrm{L}$&  $\sim 85.12$  &$7.004$~$^\mathrm{F}$\\ \hline 
         95P/(2060) Chiron&  $4.8\cross10^{18}$~$^\mathrm{B}$&  $126\cross109\cross68$~$^\mathrm{B}$& $\sim 80.94$ & $5.917813$~$^\mathrm{B}$\\ \hline 
         174P/Echeclus& $\sim 0.55 - 2.08\cross10^{17}$ &  $37.0\cross28.4\cross24.9$~$^\mathrm{P}$& $\sim 15.69 - 30.59$ &$27.785178$~$^\mathrm{R}$
    \enddata
    \tablecomments{Values with ``$\sim$'' were calculated using other values in \cref{tab:centaurs}. For Echeclus, the mass was calculated using $M = 4\pi/3 \rho  (ABC)$ with $A,\ B, \ and \ C$ from the table and $\rho$ of $500$--$1900$ kg m$^{-3}$~\citep{Pereira2024PhysicalOccultations}. Sources are indicated by superscript. L--\citet{Leiva2017}, F--\citet{Fornasier2014TheActivity}, B--\citet{BragaRibas2023Chiron}, P--\citet{Pereira2024PhysicalOccultations}, R--\citet{Rousselot2021New174P/Echeclus}.}
\end{deluxetable*}

\subsection{Centaur Activity} \label{sec:activity}

About 7-15\% of Centaurs are confirmed to be active~\citep{Bauer2008, Jewitt2024TheContinuum} with 174P/Echeclus as the most high-profile example. This activity can be periodic or transient (in short bursts with extended periods of dormancy) like Echeclus~\citep{Bauer2008, Rousselot2021New174P/Echeclus}, or regular like Chiron~\citep{Meech1990TheChiron, Luu1990CometaryChiron, Jewitt2009TheCentaurs, Dobson2024TheChiron, Belskaya2010PolarimetryChariklo, Bus20012060Aphelion} and 29P/Schwassmann–Wachmann 1~\citep{Kareta2025Activity-induced20172022, Lisse202229P/SchwassmannWachmannComets}. Chiron itself was inactive for a few years but maintained a coma~\citep{Ortiz2015PossibleChiron, Pinilla-Alonso2024UnveilingTelescope, Sickafoose2023MaterialOccultation}. Therefore, it is possible that currently dormant Centaurs were active in the past, but were not observed~\citep{Jewitt2024TheContinuum}.

The driving mechanism for Centaur activity is not well understood. It can vary from outgassing and volatile sublimation~\citep{Jewitt2009TheCentaurs} to mass-wasting events like avalanches and landslides~\citep{Brisset2025LaboratoryCentaurs, Steckloff2023ThatsAbstracts, ShiX2024, Wesoowski2020CometaryPhenomenon, Brunetti2023LandslidesSystem, Muller2024DecipheringMechanisms}. Over the course of their lifetime, icy Centaur surfaces likely go through many weathering processes that drive the surface material instability, modify surface chemistry, cause structural failure, and eventual activity~\citep{Brisset2025LaboratoryCentaurs}. Close encounters in the past with the giant planets can lead to semi-major axis ($a$) and perihelion ($q$) drops that may also trigger Centaur activity~\citep{Lilly2024Semimajor-axisComets}. Chiron is expected to have undergone an $a-$drop about $850$~yr ago~\citep[table 1 in ][]{Lilly2024Semimajor-axisComets}.
From observations, regolith is typically ejected with velocities that range on the order of $10^{-1}$ -- $10^{2}$~m s$^{-1}$ with the exact velocities depending on particle size and solar distance~\citep{ShiX2024, Meech1990TheChiron, Kareta2019Physical174P/Echeclus, Bauer2008}. Centaurs also have escape velocities on the order of 10's of m s$^{-1}$ (\cref{tab:centaurs}). 
Therefore, while the exact mechanism that drives activity on Centaurs is not well understood, activity results in the mobilization and ejection of surface regolith allowing for a variety of dynamics. 
Using these results, we explore whether mass wasting events on Centaurs can mobilize large enough quantities of regolith with enough velocity to form the precursors to the observed rings.


\section{Methods} \label{sec:methods}

Simulations of ejected regolith particles were run on Swiftest~\citep{Wishard2023Swiftest:Systems} with the RMVS integrator~\citep{Levison1994TheComets} for test particle runs and SyMBA~\citep{Duncan1998AENCOUNTERS} for massive particle runs. Both of these integrators can handle test particle-massive particle (tp-pl) and massive particle-massive particle (pl-pl) close interactions, including collisions. Test particles have no mass and do not interact with each other, i.e., no tp-tp interactions. They act as tracers of the gravity field without any additional effects. In this work, we carry out simulations where all regolith particles are either only test or only massive particles. Collisions between massive bodies can lead to either merger or fragmentation, which reduces their orbital energy while conserving momentum. The in-built collisional fragmentation model Fraggle~\citep{Wishard2023Swiftest:Systems} based on \citet{Leinhardt2012COLLISIONSLAWS} was used along with the Gravitational Harmonics routine described below in section~\ref{sec:shgrav}.

A regolith particle is considered to fall onto the Centaur at an $R_{min}$ distance and is removed from the simulation, with the mass and angular momentum added to the Centaur. We set $R_{min} = 1$~R$_{{Centaur}}$, which is the volume equivalent radius of the Centaur in our simulations. Particles that reach a distance greater than $R_{max}=50$~A$_{Centaur}$ (where A$_{Centaur}$ is the longest semi-axis of the Centaur) are also removed from the simulation. We chose this distance limit because particles that reach this far and beyond are unlikely to contribute material to the rings (current ring radii $\sim 3.1$ -- $3.3$~R$_{Chariklo}$; see \cref{tab:ring parameters}) or are on highly eccentric or escaping trajectories. All simulations were run for $100$~T$_{{rot}}$ (where T$_{rot}$ is the rotation period of the Centaur) because, as we will show later, the loss-rate of simulated regolith particles is relatively small by $100$~T$_{rot}$. The number of test particles or total mass of massive particles remaining at the end of the simulation is used to compute the capture efficiency.

\subsection{Gravitational Harmonics Routine} \label{sec:shgrav}

The two Centaurs in our study are ellipsoidal in shape~\citep{Leiva2017, Cikota2017ActivityImpacts}. This leads to a notable contribution from higher order gravitational harmonics terms, rendering a non-keplerian gravity field. In addition, most of the rings are observed to be about the 3/1 spin-orbit resonance~\citep{Sicardy2021RingsKey}, which serves as an important clue in understanding these systems. Simulations of small body rings have typically been done with spherical bodies~\citep{Michikoshi2017SimulatingChariklo, Sickafoose2024NumericalPerturber}, with only lower degree, typically zonal, terms ($m = 0$, azimuthally symmetric) like $J_2 = -C_{2, 0}$~\citep{Rimlinger2017LongRings, PanWu2016}, or only $J_2$ and $C_{2, 2}$ (second-order second-degree) terms~\citep{Winter2023OnRings}. These simulations do not encapsulate the full dynamical impact of the ellipsoidal gravity field because of the lack of higher order terms and the tesseral/azimuthal terms ($m \neq 0$), which allow for angular momentum transfer between the Centaur and ring particles~\citep{Salo2021ResonanceRings}. This effect is discussed further in section~\ref{sec:shgrav results}. 

We define $U$ as the gravitational potential at a point $\mathbf{r}$ as

\begin{equation}
\label{eqn:grav_pot}
    U(\mathbf{r}) = \frac{GM}{r} \sum_{l=0}^{\infty} \sum_{m=-l}^{l} \left( \frac{R_{\mathrm{Centaur}}}{r} \right)^l C_{lm} Y_{lm} (\theta, \phi) 
\end{equation}
where $\theta$ is the polar angle; $\phi$ is the azimuthal angle; $R_{\mathrm{Centaur}}$ is the central body radius; $M$ is the central body mass; $G$ is the gravitational constant; $Y_{lm}$ is the spherical harmonic function at degree $l$ and order $m$; $C_{lm}$ is the corresponding coefficient~\citep{Wieczorek2015GravityPlanets}. The coefficients for the Centaur are calculated by SHTOOLS~\citep{Wieczorek2018SHTools:Harmonics} using the axial measurements in~\cref{tab:centaurs}, assuming constant density, and are $4\pi$ normalized. The values of the coefficients are summarised in~\cref{tab:shgrav coefficients}. Unless mentioned otherwise, the maximum degree ($l$) of the central body Centaur is set to 6 in our simulations. This threshold was found by setting the wavelength ($\lambda = 2\pi R_{\mathrm{Centaur}} /\sqrt{l(l+1)}$) associated with a particular degree equal to the radius of the body ($R_{\mathrm{Centaur}}$)~\citep{Wieczorek2015GravityPlanets}. This threshold can be changed in Swiftest by the user if higher accuracy is required. The potential and corresponding acceleration kick is computed with these coefficients.

\begin{deluxetable}{|c|c|c|c||c|c|c|} 
    \tablecaption{The non-zero raw gravitational harmonics coefficients ($C_{l,m}$) calculated by SHTOOLS for each Centaur.\label{tab:shgrav coefficients}}
    \tablehead{\nocolhead{} & \multicolumn{3}{c||}{(10199) Chariklo} & \multicolumn{3}{c}{95P/(2060) Chiron}}
    \startdata
    \backslashbox{$m$}{$l$} & $2$ & $4$ & $6$ & $2$ & $4$ & $6$ \\ \hline
    $0$ &  $-8.630\text{E-}2$  & \phn{} $2.715\text{E-}2$ & $-1.283\text{E-}2$ & $-8.668\text{E-}2$ &  \phn{} $2.758\text{E-}2$  & $-1.325\text{E-}2$\\ \hline
    $2$ & \phn{} $2.730\text{E-}2$  & $-1.091\text{E-}2$ & \phn{} $6.502\text{E-}3$ & \phn{} $3.241\text{E-}2$ &  $-1.302\text{E-}2$  & \phn{} $7.815\text{E-}3$ \\ \hline
    $4$ &  --  & \phn{} $2.630\text{E-}3$ & $-1.597\text{E-}3$ & -- & \phn{} $3.708\text{E-}3$ & $-2.262\text{E-}3$\\ \hline
    $6$ &  --  & -- & \phn{} $3.908\text{E-}4$ & -- & -- & \phn{} $6.504\text{E-}4$ \\
    \enddata
    \tablecomments{The coefficient values for the odd degree/order and $m < 0$ terms are $0$ because of symmetry in the assumed perfect ellipsoid shape models. The $l = 0$ term is accounted for in the Keplerian dynamics because it refers to the point source or perfectly spherical component. The coefficients are $4 \pi$ normalized and not scaled by the body radius.}
\end{deluxetable}

We added a routine to Swiftest that calculates the acceleration kick $\bm{\alpha(r)} = \bm{\nabla} U(\mathbf{r})$~\citep{Wieczorek2015GravityPlanets} with the gravitational harmonics where $U$ is the potential in~\cref{eqn:grav_pot}. This generalized routine captures the full effect of the central body, including the tesseral terms. The azimuthal terms (such as $C_{2,2}$) are shown to help in the relaxation and formation of ring structure~\citep{Winter2023OnRings}. We see in section \ref{sec:shgrav results} that the gravitational harmonics play a significant role in capture and dynamics of ejected regolith particles. They can excite regolith particles and clear them out or restrict them to a region. 

\begin{figure}[h]
    \centering
    \plotone{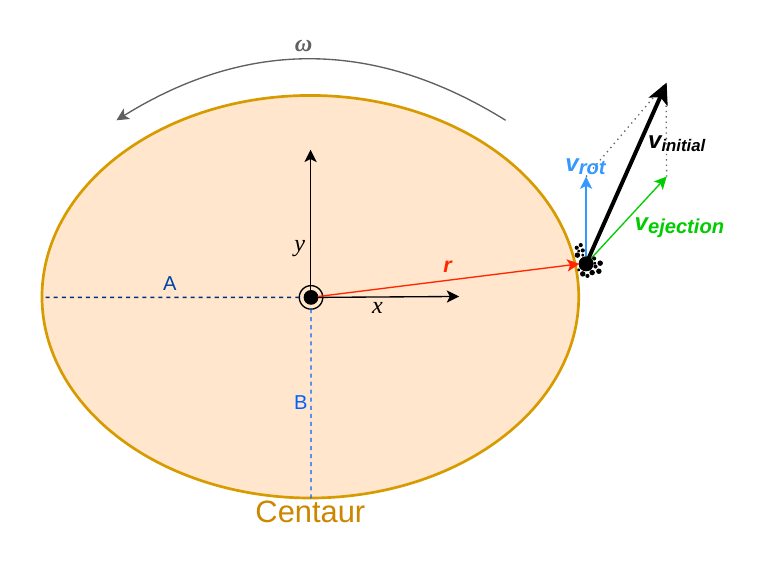}
    \caption{Initial simulation setup in the inertial reference frame of Swiftest~\citep{Wishard2023Swiftest:Systems}. Here we show regolith ejection from the long axis in the equatorial plane. The orange ellipsoid is the equatorial plane of the Centaur when seen top-down (x--y~plane, z-axis out of the page). Regolith particles, represented by the black dots on the right, are ejected from the surface of the Centaur in the equatorial plane with $\mathbf{r}$ being the position vector. The blue arrow ($\mathbf{v}_{rot}$) shows the centrifugal correction from the non-inertial rotating frame of the Centaur. The green arrow ($\mathbf{v}_{ejection}$) is the ejection velocity of the regolith that we test and vary. The black arrow ($\mathbf{v}_{initial}$) is the net velocity vector of the regolith. The Centaur's rotation rate denoted by $\bm{\omega}$ is assumed to be around the $\mathbf{z}$-axis in the x--y plane (out of the page).}\label{fig:ejection_setup}
\end{figure}

\begin{figure}[h]
\centering
\plotone{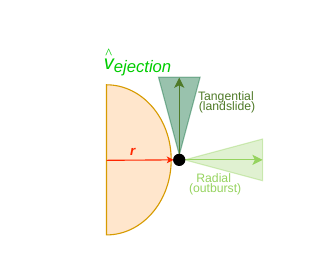}
\caption{A close-in diagram explaining the terminology for the ejection direction of $\mathbf{v}_{ejection}$. Regolith particles are ejected in either in a landslide or radial direction with an angle variance up to $30^\circ$ to add variability (indicated by the shaded regions) in a non-inertial frame defined by the Centaur's rotation. We then test capture rates for these radial-like and landslide-like ejection trajectories. Figures are not to scale. Symbols with the~$\hat{}$~symbol indicate only the direction component of that vector value without commenting on the magnitude.\label{fig:vel_direction}}
\end{figure} 

\subsection{Initial Setup and Ejection Parameters} \label{sec:simulation setup}

\Cref{fig:ejection_setup} illustrates the initial ejection setup where $500$ regolith particles are ejected in the equatorial plane (x-y plane in the simulations) of the Centaur. Their initial positions are randomized just above the Centaur's surface, along the long axis (x-axis) or short axis (y-axis) with a slight perpendicular deviation from the originating axis. They are then given an initial velocity vector ($\mathbf{v}_{initial}$) which is split into two components: $\mathbf{v}_{ejection}$ and $\mathbf{v}_{rot}$ as
\begin{equation}
    \label{eqn:init_vel}
        \mathbf{v}_{initial} = \mathbf{v}_{ejection} + \mathbf{v}_{rot}
\end{equation}
where we vary the ejection velocity ($\mathbf{v}_{ejection}$) to mimic either a landslide or a radial ejection with a randomized angle variance of up to $30^\circ$ (\cref{fig:vel_direction}) and test the capture likelihood as a function of magnitude and direction. $\mathbf{v}_{rot}$ is the centrifugal contribution, i.e., the correction to the velocity of the regolith particles as we switch from the non-inertial rotating frame of the Centaur to the inertial frame of the simulations, defined as 
\begin{equation}
    \label{eqn:v_rot}
    \mathbf{v}_{rot} = \bm{\omega} \cross \mathbf{r}
\end{equation}
where $\bm{\omega}$ is the rotation vector of the Centaur and $\mathbf{r}$ is the initial position vector of the regolith particle. The centrifugal contribution is due to the rotating Centaur and the ejected regolith particles are initialized just above the Centaur's surface. We account for the small technicality that arises from the initial positions being just above, but not on, the surface, though it does not change the results.

Ejection from the equatorial plane gives the largest $\mathbf{v}_{rot}$ contribution and allows for easier capture. We modeled the ejected regolith both as test particles and massive particles, and compared the results of the two types of simulations. Test particles do not interact with each other and act as tracers of the gravity field around the Centaur. As a result, we can constrain and understand the dynamical effect of the gravity field on each particle without any other interactions. Because the central bodies (tested Centaurs) are ellipsoidal, their gravitational harmonics are significant and perturb the Keplerian orbits of the regolith particles. Because of the non-spherical gravity field, we no longer have a simple Keplerian 2-body problem where an ejected particle can do one of two things: fall back onto the body or escape. Gravitational harmonics up to degree $l = 6$ and order $|m| \leq l$ are included. See section~\ref{sec:shgrav} for more details.

By giving the regolith particles mass, we are building off of the results of the test particle simulations by studying regolith-regolith interactions. The regolith now have inter-particle gravity, and are able to collide and fragment with each other. As we will demonstrate, these regolith-regolith interactions significantly help with orbit capture. The total mass ejected in our simulations is on the order of the current ring mass of $\sim 8 \times 10^{13}$~kg (see section \ref{sec:current rings}) with the regolith particles having the same density as the Centaur. We picked this value for initial mass so as to directly compare the required outburst size to the current expected mass in the rings. This total ejected mass is larger than the amount of mass ejected from Echeclus~\citep[$\sim 10^9$~kg,][]{Bauer2008} but there is minimal data on total mass loss from singular outburst events. Because we analyze our results in terms of the fraction of material that remains in orbit, the initial mass value does not significantly change the results. However, because of numerical constraints on our N-body model, simulated regolith particles are much larger than they would be in reality, with each individual regolith particle radii varying from $200 - 500$~m~\citep[compared to a true size of $> 1 \mu$m for Chariklo,][]{Morgado2021Refined2020}. If the regolith particles were smaller and more realistic, we expect the same large-scale result. We would expect more collisions because of the increase in number density, less fragmentation, and potentially more collisional damping for the same total mass. However, we cannot comment on specifics because the collisional fragmentation models in Swiftest are not well constrained for bodies well into the strength regime~\citep[$R \lesssim 100$~m,][]{Stewart2009Velocity-DependentPlanetesimals}. Accordingly, our final results are quantified by total mass of the regolith particles.

The central body in the simulation is the Centaur of uniform density with mass and axes values from \cref{tab:centaurs}. We test Chariklo and Chiron, with their dimensions taken from table \ref{tab:centaurs}. The units are normalized to the central body (i.e. Centaur radius for length, Centaur mass for mass, and rotation period for time) with velocity parameterized by the Centaur's escape velocity. The simulation parameters are summarized in table~\ref{tab:initial parameters}. The setup is the same for each ejection direction (with a $30^\circ$ variation) and is repeated for each regolith particle type and Centaur.

\begin{deluxetable*}{|c|c|c|}[h]
    \tablecaption{Variable model parameters used in our simulations.}\label{tab:initial parameters}
    \tablehead{\colhead{Parameter} & \colhead{Range of values tested} & \colhead{No. of Variations}}
    \startdata
    \multicolumn{3}{c}{Variable Parameters} \\ \hline
    Central body & (10199) Chariklo & 2 \\ 
    \phn & 95P/(2060) Chiron & \phn \\ \hline
    Regolith particle type & Massive particle & 2 \\
    \phn & Test particle & \phn \\ \hline
    Ejection direction & Tangential (long-axis) & 3 \\
    \phn & Tangential (short-axis)& \phn \\ 
    \phn & Radial & \phn \\ \hline
    Ejection velocity & $0 - 1.0$~v$_\mathrm{escape}$ & 10\\
    \phn & (in bins of $0.1$~v$_\mathrm{escape}$)\tablenotemark{a} & \phn \\ \hline
    Regolith radius & $200 - 500$~m & Randomized \\ \hline
    Gravitational harmonics degree and order & $l = 0 - 6; \ |m| \leq l$ & 3\tablenotemark{b} \\
    \phn & $l = 0 - 6; \ |m| = 0$ & \phn \\
    \phn & $l = 0 - 6; \ 1 \leq |m| \leq l$ & \phn \\ \hline
    \multicolumn{3}{c}{Constant Parameters} \\ \hline
    $N_\mathrm{particles}$ & \multicolumn{2}{c|}{$500$} \\ \hline
    Chariklo regolith density & \multicolumn{2}{c|}{$\sim 763.22$~kg m$^{-3}$} \\ \hline
    Chiron regolith density & \multicolumn{2}{c|}{$\sim 1227.01$~kg m$^{-3}$} \\
    \enddata
    \tablecomments{The regolith particle densities are assumed to be the same as the Centaur and are calculated using \cref{tab:centaurs}.}
    \tablenotetext{a}{We ran 1 simulation for Chariklo for a tangential ejection from the long-axis with a ejection velocity bin size of $0.3$~v$_\mathrm{escape}$ (range of $0.2 - 0.5$~v$_\mathrm{escape}$) shown in \cref{fig:sim regolith capture combined x vs y,fig:sim regolith capture combined a vs e}.}
    \tablenotetext{b}{For a given Centaur, the gravitational harmonics are only varied for the analysis in section~\ref{sec:shgrav results}. In regolith capture simulations (sections~\ref{sec:radial}~--~\ref{sec:Chiron}), the terms span $l \leq 6$ and $|m| \leq l$.}
\end{deluxetable*}

\section{Results} \label{sec:results}

We test various parameters for regolith capture with each summarized below. As mentioned above, particles are ejected with a radial (outburst) or tangential (landslide) like ejection (in addition to the $\mathbf{v}_{rot}$) across a variety of initial ejection velocity magnitudes. First we test ideal ejection trajectories, initial ejection velocities, location of ejection, and gravitational harmonics terms. Then, we quantify the amount of regolith mass in orbit around the Centaur versus time for various initial $v_{ejection}$ values, and the amount of regolith mass captured at the end of our simulations ($t = 100$~T$_{{rot}}$) versus initial ejection velocity bin ($v_{ejection}$). For our purposes, ``capture" is defined as the amount of regolith in an orbit around the Centaur with eccentricity $e < 1$ and periapsis $q > 1$~R$_{Centaur}$. For the test particle cases, ``mass" refers to the number of particles. The simulations below in sections~\ref{sec:radial}~--~\ref{sec:longitude} were performed on Chariklo. Simulations with Chiron in section~\ref{sec:Chiron} showed similar results.

\begin{figure*}[h!]
\centering
    \plotone{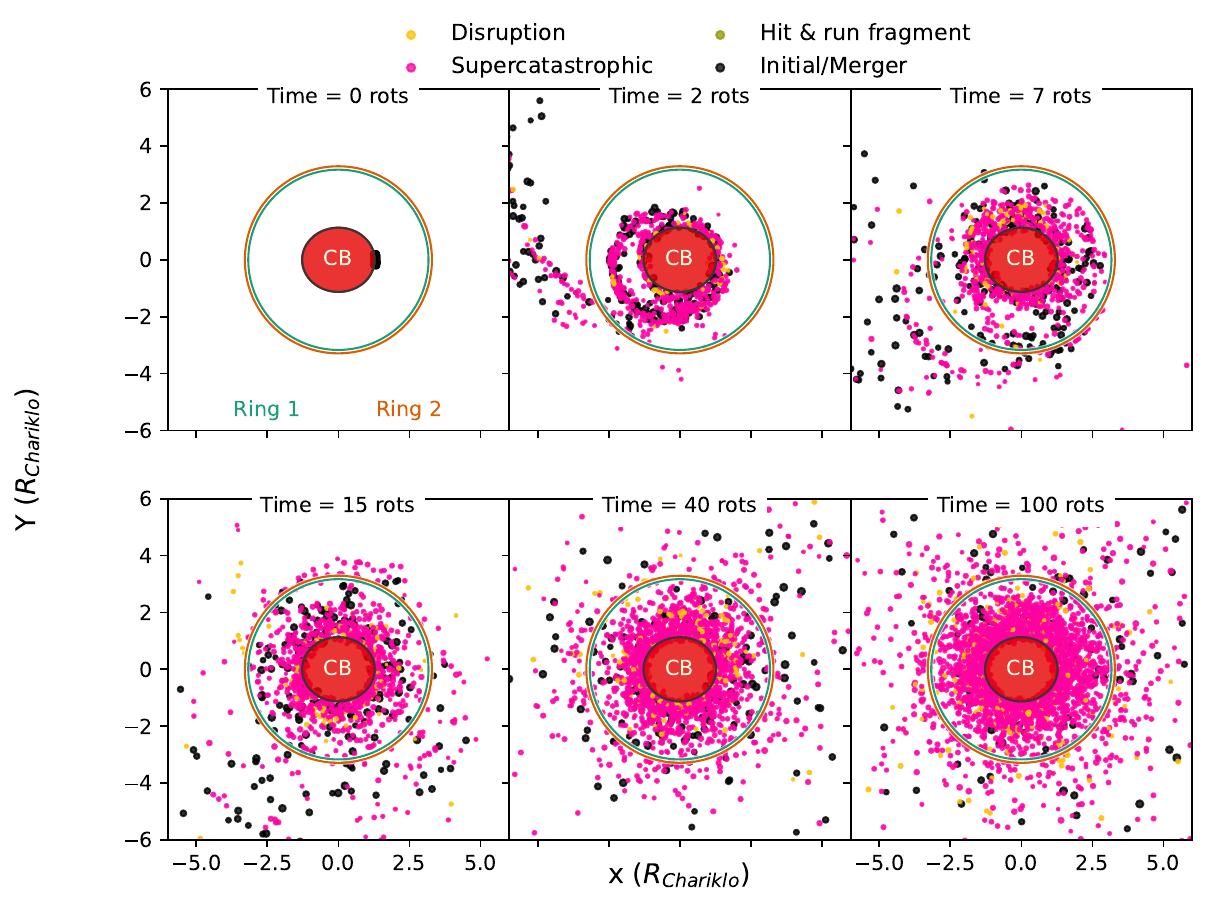}
    \caption{The x vs y progression of regolith ejection and capture in the equatorial plane around Chariklo (marked as Central Body (CB)). We show a landslide-like ejection from the long axis with the initial $v_{ejection} = 0.2 - 0.5$~v$_\mathrm{escape}$. We start with $500$ regolith particles and end with $3657$ regolith particles in this simulation. While a large number of regolith particles are captured, some fall back onto the Centaur or escape. We start to see stable orbits in $1-3$~rotations and the regolith particles (visually) form an initial disk-like state in $\sim15$~rotations. Particles are color-coded by collisional origin with their size scaled up for clarity. \\ NOTE--There are no hit-and-run fragments generated in these simulations. Regardless, we have included that collisional category in the legend for completeness.}\label{fig:sim regolith capture combined x vs y}
\end{figure*}

\begin{figure*}[h!]
\centering
    \plotone{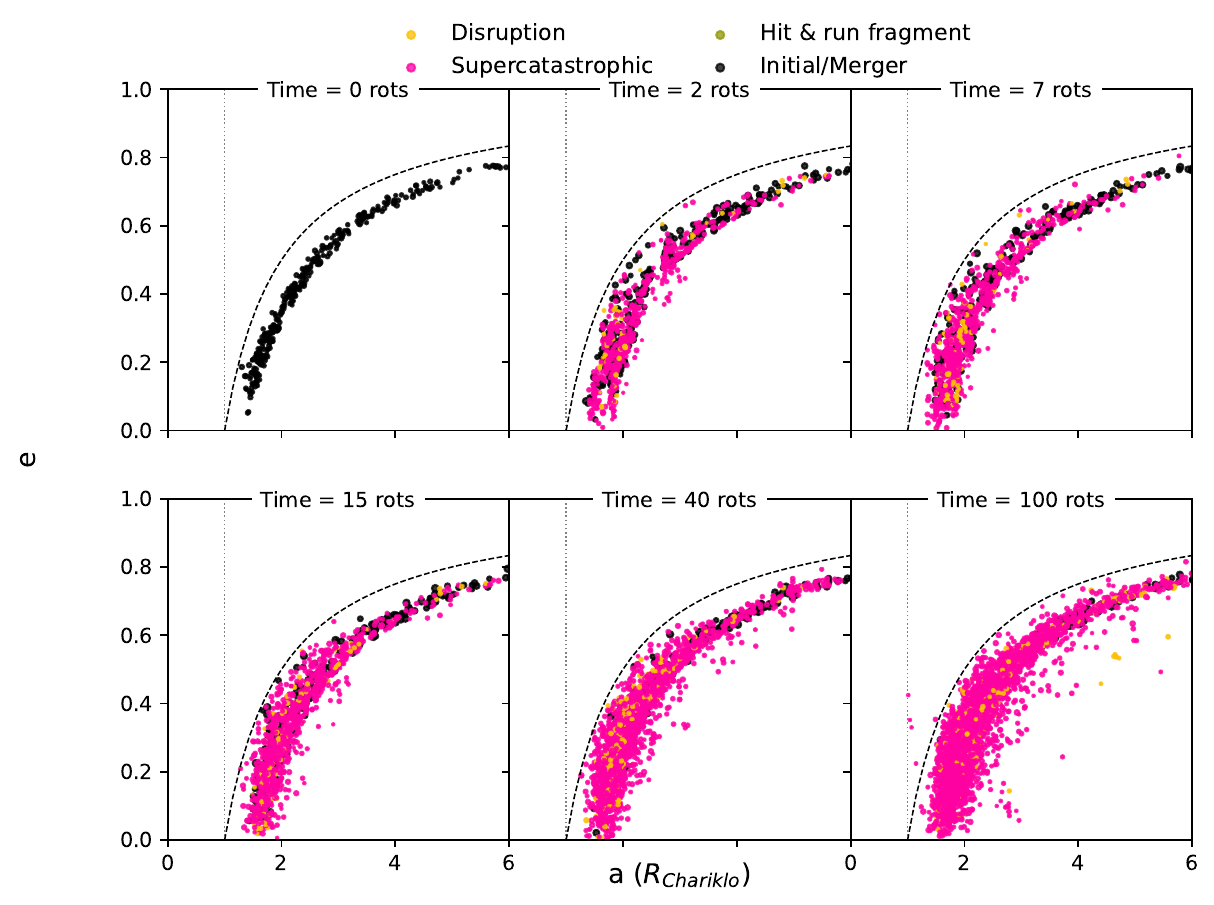}
    \caption{Semi-major axis vs eccentricity ($a$ vs $e$) plot of regolith capture of the same simulation in \cref{fig:sim regolith capture combined x vs y}. We show a landslide-like ejection from the long axis with the initial $v_{ejection} = 0.2 - 0.5$~v$_\mathrm{escape}$. The black dashed line is the pericenter curve $q = 1$ $R_{Chariklo}$ and the gray dotted line is $a = 1$ $R_{Chariklo}$. The regolith particles are dynamically excited and bounce back and forth along the pericenter curve because of the ellipsoidal Centaur. Collisions between particles damp their energy and push them further under the pericenter curve.}\label{fig:sim regolith capture combined a vs e}
\end{figure*}

The first case to check for regolith capture is if ejected regolith particles can be captured solely by regolith-Centaur (gravity) and then with regolith-regolith interactions (gravity and collisions). Swiftest simulation freeze-frames in \cref{fig:sim regolith capture combined x vs y,fig:sim regolith capture combined a vs e} show the typical ejection and capture process for a landslide-like ejection of massive particle regoliths. Regolith particles are captured in a proto-ring disk. As a reminder, the gravitational harmonics terms in the simulations in sections~\ref{sec:radial} -- \ref{sec:Chiron} span $l \leq 6$ and $|m| \leq l$.

Massive particle simulation results are color coded by collisional origin, i.e., the type of fragmenting collision that created that particle, of which Swiftest has four. A ``Supercatastrophic" disruption collision is when the target body loses $\geq  50 \%$ of its mass while a ``Disruption" collision occurs when the target body loses $< 50\%$ of its mass. A ``Hit \& run" fragment is generated from a grazing impact with a high impact parameter and an ``Initial/Merger" fragment is an initial condition particle that may have accumulated more mass via mergers. The different types of collisions are defined in more detail in~\citet{Leinhardt2012COLLISIONSLAWS}. The specific collisional origin is impertinent for ring capture behavior as disruptive collisions in general help with orbit capture by losing energy and damping excited particles. We see more ``supercatastrophic" fragments in~\cref{fig:sim regolith capture combined x vs y,fig:sim regolith capture combined a vs e} than others because, by definition, these collisions involve complete breakup of the target body and thus, create a lot of fragments. This effect is further amplified because of the large regolith particles that we have chosen due to computational constraints. Regardless, this definitively implies that the initial velocity dispersion between particles is large and a lot of fragmenting collisions can occur. The underlying mechanisms and the expectation that inter-particle collisional damping aids regolith capture stay the same.

\subsection{Radial Ejection (from the long axis)} \label{sec:radial}

\begin{figure}[h!] 
\centering
\epsscale{0.8}
\plotone{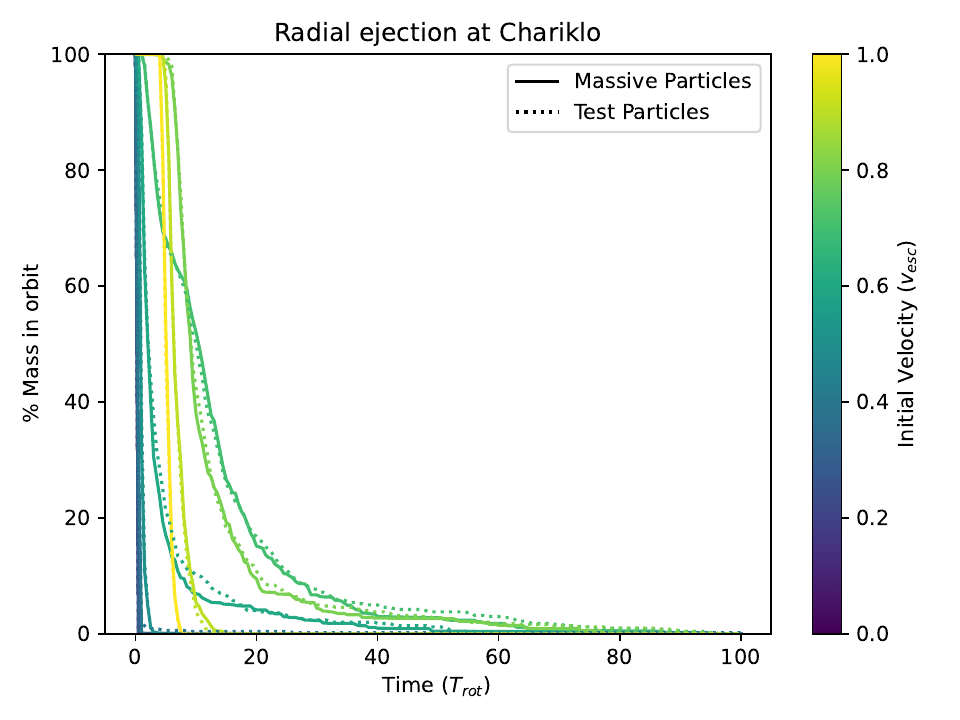}
\caption{Regolith mass in orbit vs time for a radial ejection around Chariklo from the long axis. Test particle regoliths are marked by the dotted lines and massive particle regoliths by the solid lines. Regolith particles are lost from the system extremely quickly across all velocity ranges, regardless of particle type, showing that radial ejections are not favorable for regolith capture. Simulations are parameterized by the initial $v_{ejection}$.\label{fig:radial mass capture}}
\end{figure}

We launch regolith particles on a radial-like trajectory from the longest axis in the equatorial plane, and see effectively no capture for both test and massive particle cases (figure \ref{fig:radial mass capture}). However, some massive particle radial-like ejections do show small amounts of capture ($<0.5\%$) on very big orbits ($a > 10$~R$_{\mathrm{Centaur}}$). From \cref{tab:centaurs}, the rings are at $\sim 3.2 - 3.3$~R$_\mathrm{Centaur}$, and this makes a radial ejection unfavorable for proto-ring material capture. 

\subsection{Landslide Ejection (from the long axis)} \label{sec:landslide}

Next, we launch regolith particles in a landslide-like ejection from the long-axis in the equatorial plane as seen in~\cref{fig:sim regolith capture combined x vs y,fig:sim regolith capture combined a vs e}. Here we find that a landslide-like ejection is ideal for regolith capture in the absence of a debris disk. An initial ejection velocity in the $0.2 - 0.5$~v$_{\mathrm{escape}}$ range shows maximum capture when regolith particles are ejected from the long axis in the equatorial plane (see the left panel of figure \ref{fig:chariklo landslide mass capture combined}). Comparing that with figure \ref{fig:radial mass capture}, we see that regolith particles ejected in the right velocity range can be temporarily caught in orbit around the body, corroborating \citet{Laipert2014SatelliteKleopatra}. Test particle regoliths (dotted lines in \cref{fig:chariklo landslide mass capture combined}) show quasi-stable orbits for the duration of our simulations, but they trend towards total regolith particle loss. Massive particle regoliths show clear orbital stability and a significant increase ($\sim 7 - 15 \times$) in mass capture as compared to the test particle case. \Cref{fig:chariklo landslide hist combined} compares the amount of regolith mass captured at $t = 100$~T$_{rot}$ for each ejection velocity bin between the test particle and massive particle cases.

\subsection{Landslide Ejection (from the short axis)} \label{sec:longitude}

\begin{figure}[h]
\centering
\plottwo{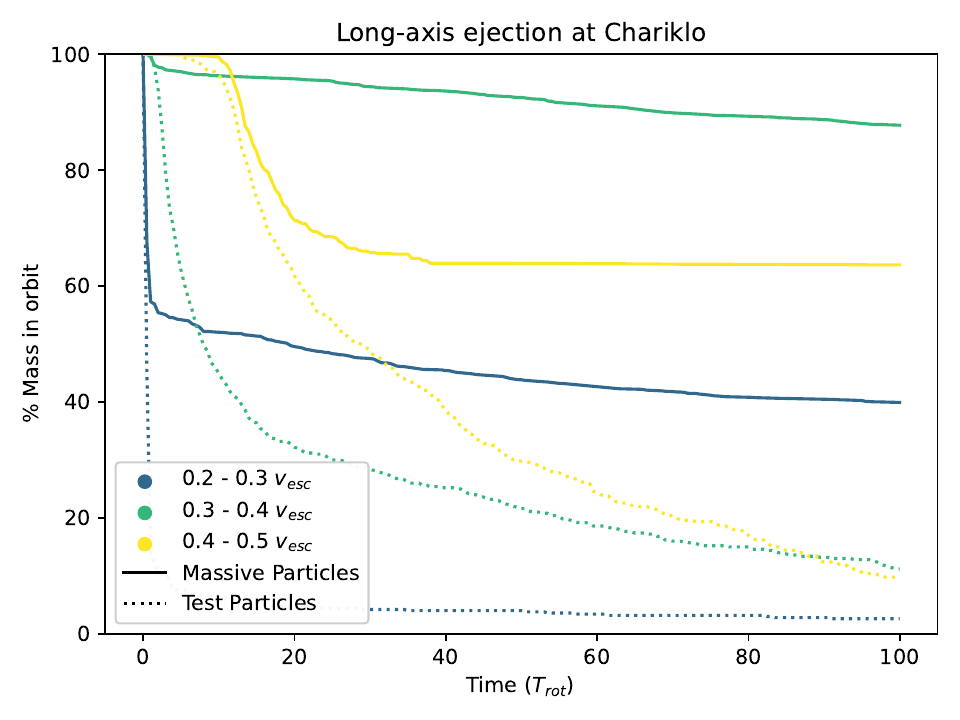}{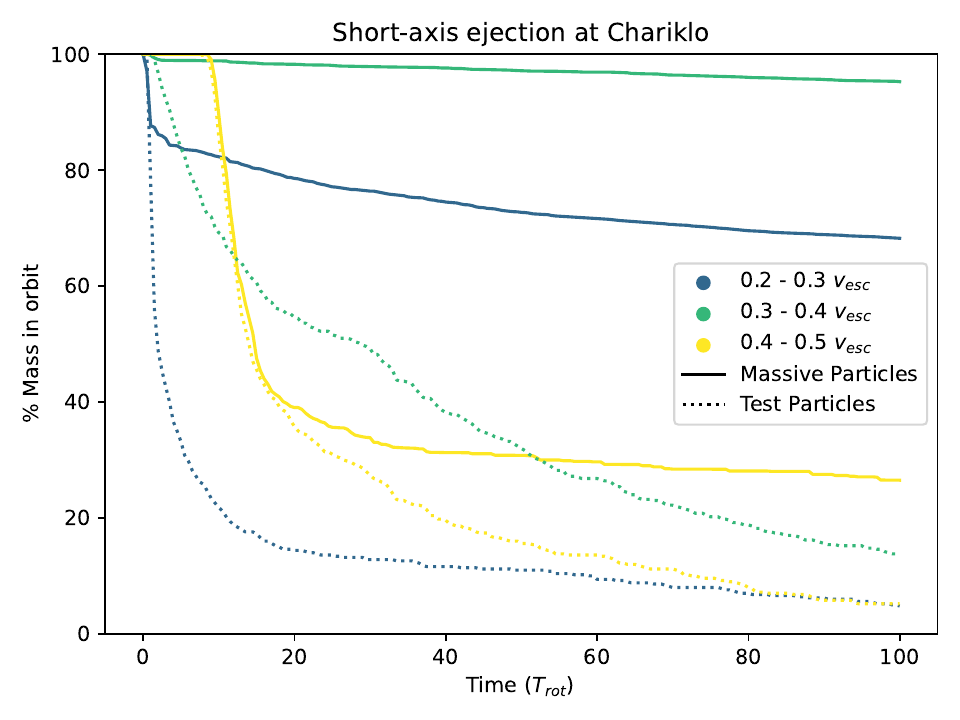}
\caption{Regolith mass in orbit vs time for a landslide-like ejection from the long-axis (left panel) and short axis (right panel) around Chariklo. Test particle regoliths are marked by the dotted lines and massive particle regoliths by the solid lines. Regolith particles are lost very quickly in the low and high velocity ranges (not shown for clarity), but stay in orbit for an extended period of time for some velocities. Massive particle interactions and collisions help counter the angular momentum transfer from the ellipsoidal Centaur and help keep regolith particles in orbit over the course of our simulations. Simulations are parameterized by the initial $v_{ejection}$.\label{fig:chariklo landslide mass capture combined}}
\end{figure}

\begin{figure}
    \centering
    \plottwo{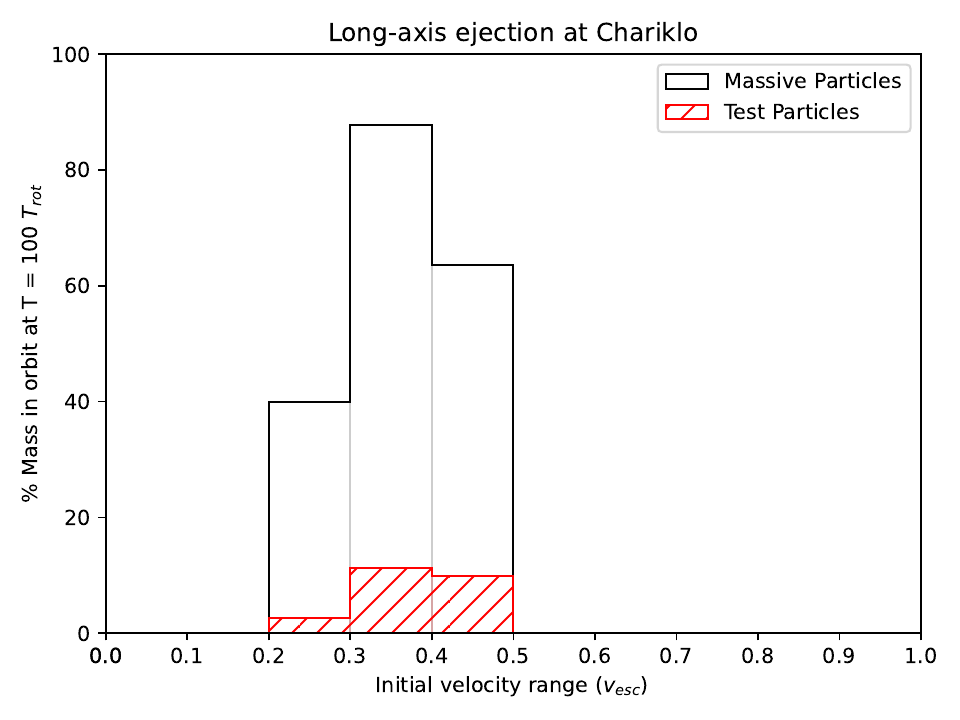}{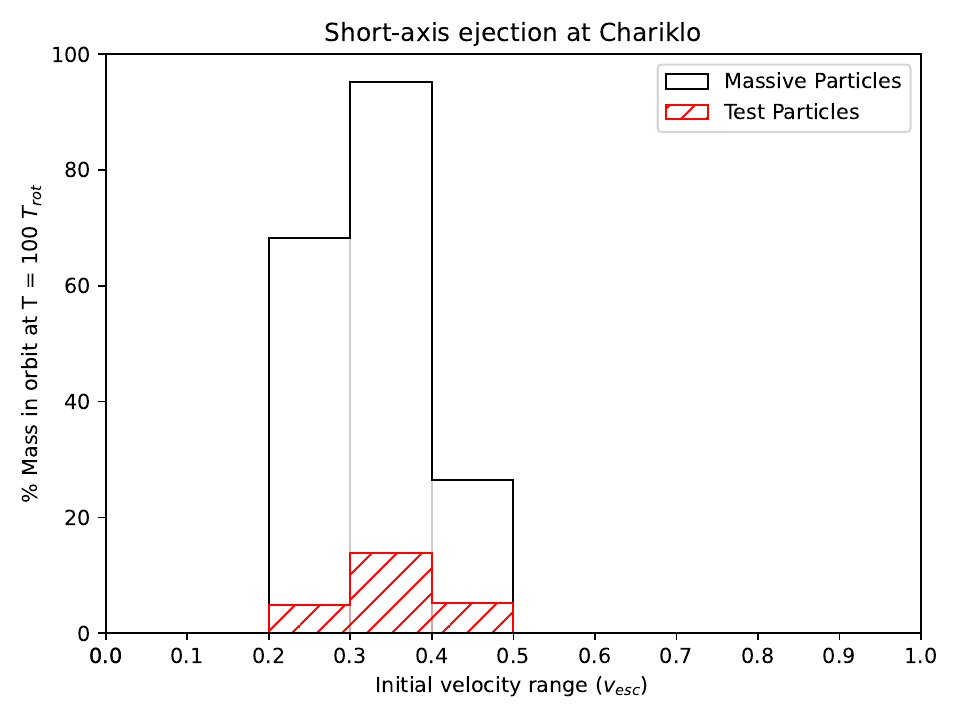}
    \caption{Mass in orbit at $t = 100$~T$_\mathrm{rot}$ per initial $v_{ejection}$ bin for a landslide ejection from the long-axis (left panel) and short axis (right panel). Test particle regoliths are marked by the red hatches and massive particle regoliths by the black solid lines. Here we can see the amount of regolith captured in each initial velocity bin at the end of our simulations. We see high levels of capture up to $\sim 95 \%$ in the massive particle case in the right panel.\label{fig:chariklo landslide hist combined}}
\end{figure}

Building off the previous section, we eject regolith particles from the short axis in the equatorial plane with a landslide-like ejection. This means that the regolith particles have a lower centrifugal velocity kick and are off-phase with the longest axis. In the right panels of figures \ref{fig:chariklo landslide mass capture combined} and \ref{fig:chariklo landslide hist combined}, we plot the captured mass for these simulations. Qualitatively similar to section \ref{sec:landslide}, we see higher overall capture and a favorable shift to a lower $v_{ejection}$ range in this case, as seen in figure \ref{fig:chariklo landslide hist combined}. This implies that ejections from the longest axis favor a higher $v_{ejection}$ while ejections from the short axis favor a lower $v_{ejection}$ for orbit capture.

Therefore, landslide-like ejections are favorable for regolith orbit capture. The initial Keplerian orbit places them close to the pericenter curve where a combination of collisions and the ellipsoidal gravity field aid orbit capture. Radial-like ejections place regolith particles on very eccentric initial orbits that lead to eventual escape or a very large semi-major axis. The mass capture rates are summarized in table \ref{tab:mass capture} with typical ring capture with massive particles shown in \cref{fig:sim regolith capture combined x vs y,fig:sim regolith capture combined a vs e}.

\subsection{Chiron} \label{sec:Chiron}

\begin{figure}[h]
    \centering
    \plottwo{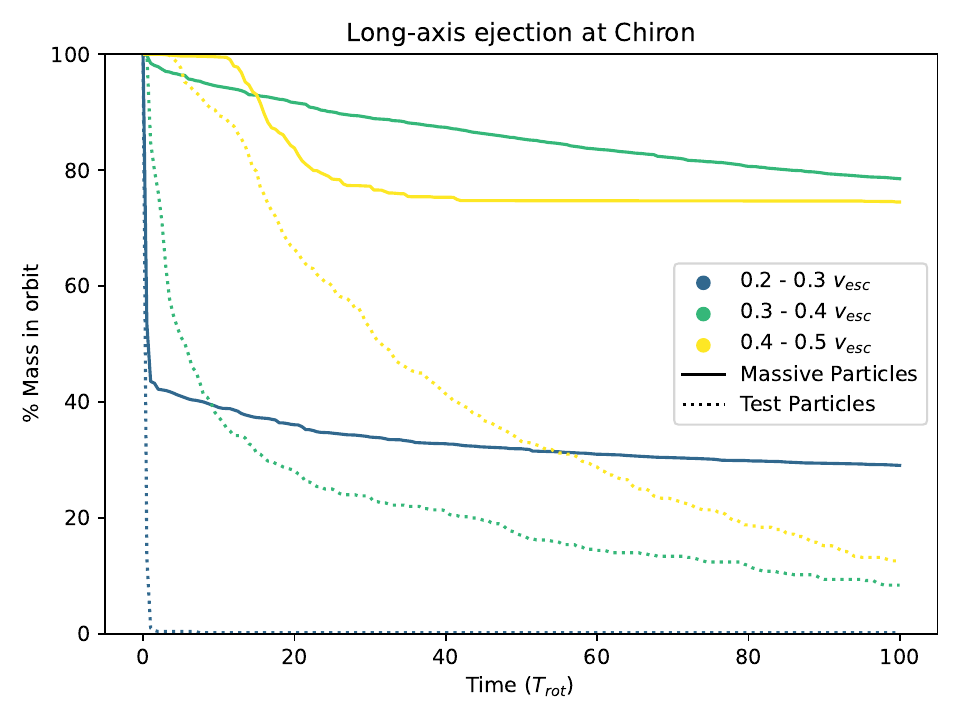}{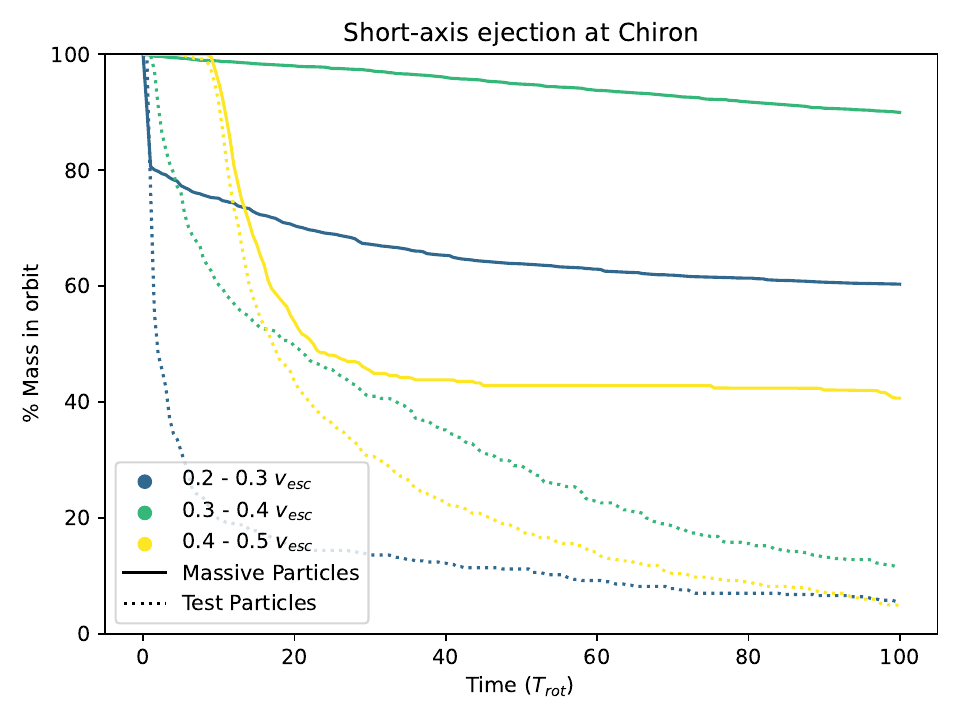}
    \caption{Regolith mass in orbit vs time for a landslide-like ejection around Chiron. Here we only show results from the velocity ranges that show significant capture for clarity. Ejection from the long-axis is on the left-hand side and short-axis on the right-hand side. Regolith capture around Chiron shows qualitatively similar results to Chariklo because the gravitational harmonics are similar when normalized. Simulations are parametrized by the initial $v_{ejection}$.\label{fig:chiron landslide mass capture}}
\end{figure}

\begin{figure}[h]
    \centering
    \plottwo{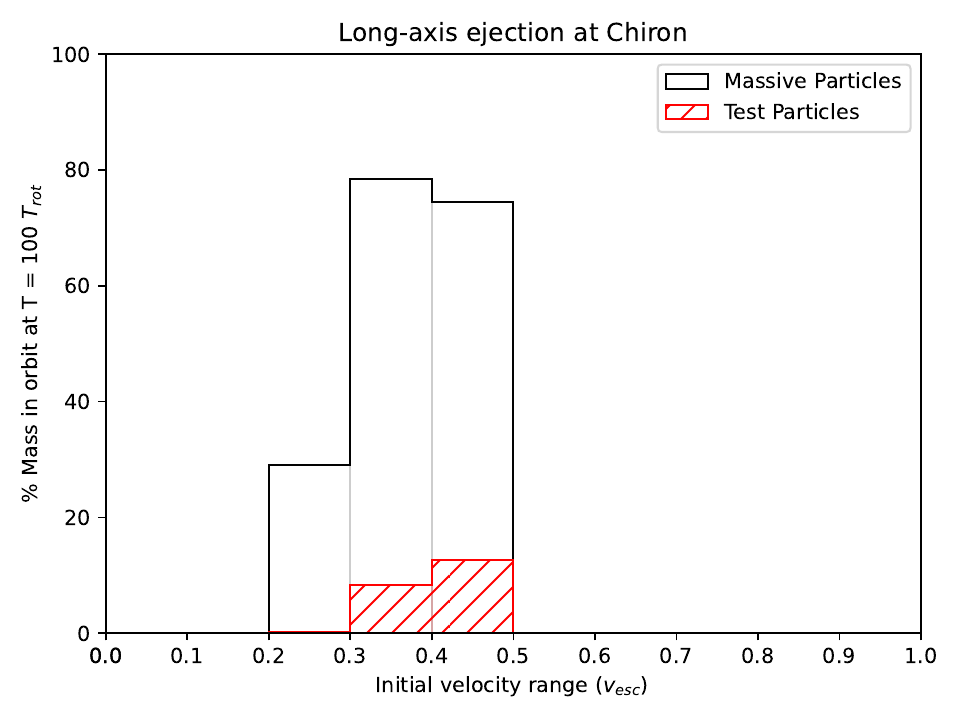}{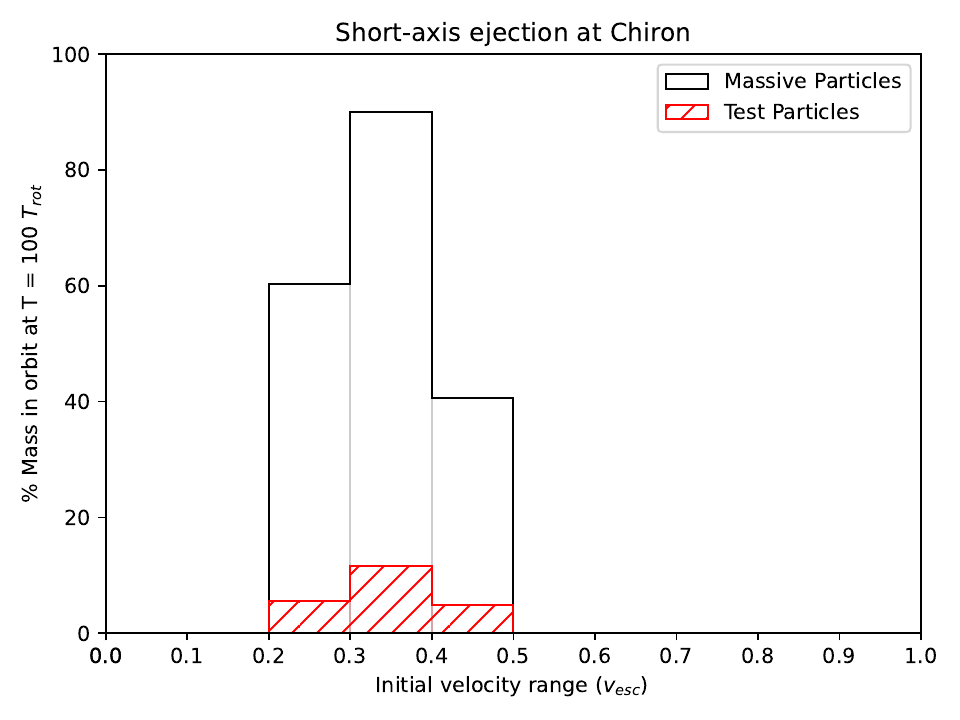}
    \caption{Mass captured at $t = 100$~T$_\mathrm{rot}$ per initial $v_{ejection}$ bin for a landslide ejection around Chiron. Here we only show results from the velocity ranges that show significant capture for clarity. Regolith ejection from the long-axis is on the left-hand side and short-axis on the right-hand side. Regolith capture behavior for Chiron is similar to Chariklo qualitatively, but differs a bit quantitatively.\label{fig:chiron landslide hist}}
\end{figure}

Chiron is smaller but similar to Chariklo in physical characteristics when taking the triaxial ellipsoidal shape model (\cref{tab:centaurs}). We ran simulations for regolith capture on Chiron and saw similar results as with Chariklo. Radial-like ejections showed no capture while landslide-like ejections from the long and short axis on the equatorial plane showed capture as seen in \cref{fig:chiron landslide mass capture} and \ref{fig:chiron landslide hist}. Capture amounts are summarized in \cref{tab:mass capture} for both Chariklo and Chiron.

Looking at the nature of regolith capture in our simulations (\cref{fig:sim regolith capture combined x vs y,fig:sim regolith capture combined a vs e}), Chiron serves as the closest real-life example to our work. The regolith in caught in orbit as a disk, somewhat similar to the debris observed around Chiron~\citep{Ortiz2015PossibleChiron, Ortiz2023The15, Pereira2025TheSystem, Sickafoose2020Characterization2011,Sickafoose2023MaterialOccultation}. Over the course of our relatively short simulations, we do not create restricted bands of dense material currently observed around Chiron. While the tesseral gravitational harmonics terms~\citep{Winter2023OnRings} and/or spin-orbit resonances~\citep{Sicardy2021RingsKey} could be potential avenues to create confined structures of orbiting material, this does not happen for excited disk particles on timescales shorter than $100$~rotations. 

\begin{deluxetable}{|c|c|c|c|c|}[h]
     \tablecaption{Percentage of regolith in orbit at $100$~T$_{rot}$ around a Centaur for a variety of landslide ejections.\label{tab:mass capture}}
    \tablehead{\colhead{Initial $v_{ejection}$ (v$_{\mathrm{escape}}$)} & \multicolumn{2}{|c}{Long Axis} & \multicolumn{2}{|c}{Short Axis} \\ \hline
    \nocolhead{} & \multicolumn{1}{|c}{Chariklo} & \multicolumn{1}{|c}{Chiron} & \multicolumn{1}{|c}{Chariklo} & \multicolumn{1}{|c}{Chiron}}
    \startdata
    $0.2 - 0.3$& $39.93\% \ (2.6\%)$ & $29.06 \% \ (0.2 \%)$ & $68.25\% \ (4.8\%)$ & $60.33 \% \ (5.6 \%)$ \\ \hline 
    $0.3 - 0.4$&  $87.75\% \ (11.2\%)$& $78.52 \% \ (8.4 \%)$ & $95.25\% \ (13.8\%)$ & $89.96\% \ (11.6\%)$\\ \hline 
    $0.4 - 0.5$& $63.66\% \ (9.8 \%)$ & $74.51\% \ (12.6 \%)$ & $26.51\% \ (5.2 \%)$ & $40.68\% \ (4.8 \%)$ \\
    \enddata
    \tablecomments{Massive particle regolith results shown with test particle results in parentheses.}
\end{deluxetable}

\subsection{Capture Mechanism} \label{sec:capture mechanism}

\begin{figure}[h]
    \centering
    \epsscale{0.8}
    \plotone{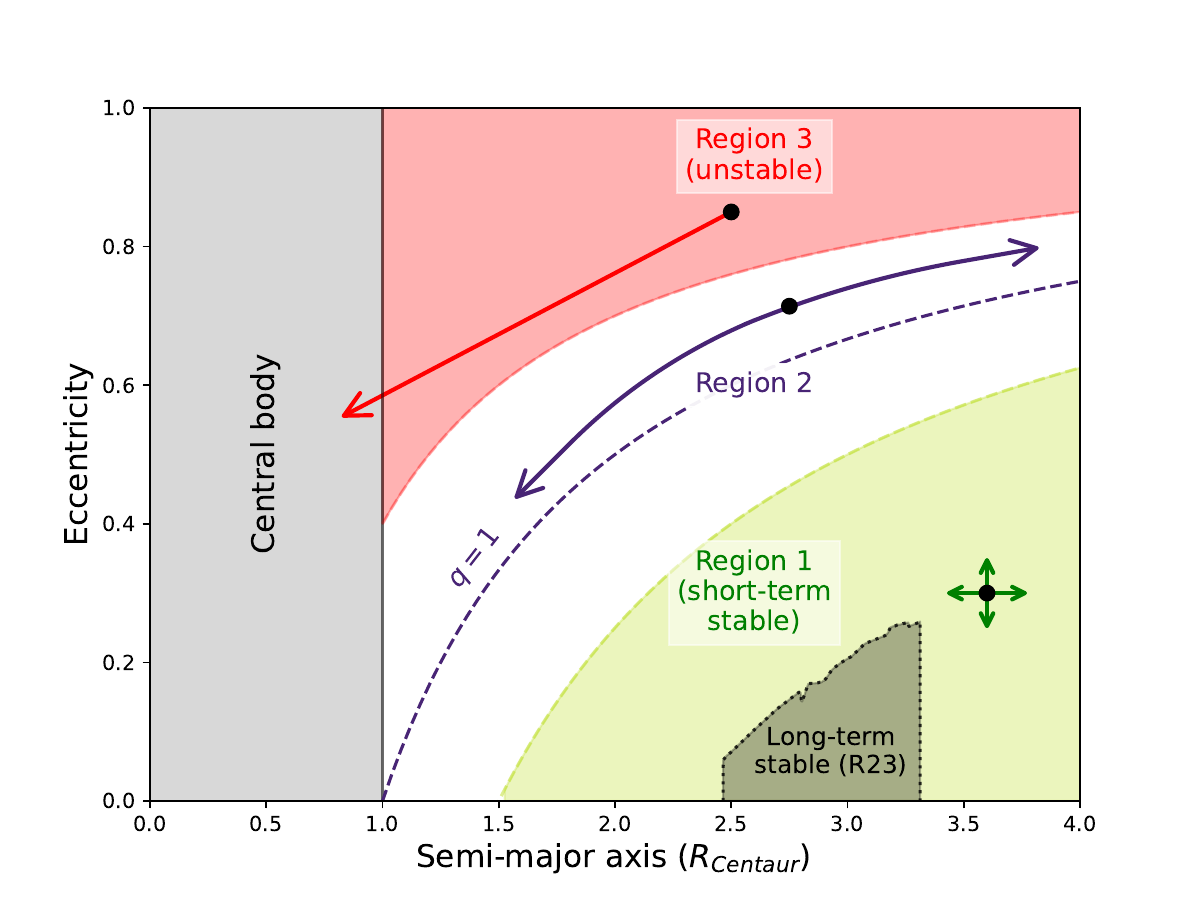}
    \caption{This cartoon qualitatively explains the general process by which the ellipsoidal Centaur alters the ejected regolith (black dot) orbits in $a-e$ space. The purple dashed line denotes an orbit with a pericenter of $q = 1$~R$_\mathrm{Centaur}$. Regolith particles with orbits of $q \leq 1$~R$_\mathrm{Centaur}$ or $ a \leq 1$~R$_\mathrm{Centaur}$ (gray region) crash into the Centaur and are removed from the system.
    Region 1 (green) is our short-term stable region with $q > 1$~R$_\mathrm{Centaur}$ where particles hover around their initial orbit (upto $100$~orbits). We have overlaid the more restricted and long-term stable region (upto $\sim 10,000$~orbits) around Chariklo from~\citet{Ribeiro2023DynamicsBody}. Region 2 (white between the green and red regions) is a quasi-stable region with particles shuttling back and forth due to angular momentum transfer from the ellipsoidal Centaur. Region 3 (red) is an unstable region with $q < 1$~R$_\mathrm{Centaur}$ where particles are pulled in towards, and eventually crash into the Centaur. Ejected regolith particles typically have initial orbits with high $e$ and low $a$, near Region 2 and 3. The ellipsoidal Centaur initially ``pulls down" regolith close to/under the $q = 1$~R$_\mathrm{Centaur}$ curve. Regolith particles are brought closer together, increasing the likelihood of collisions. Subsequent collisions between regolith particles also damp their orbits, transfer them into the stable regions (region 1 and 2), and aid capture into orbit. All lines drawn are approximate.}
    \label{fig:clm effect} 
\end{figure}

The gravity field of the ellipsoidal fast rotating Centaur is the main reason for this orbit capture. The higher order gravitational harmonics terms significantly alter the initial Keplerian orbit and the azimuthal terms allow for consistent angular momentum transfer between the Centaur and regolith particles. This capture is best understood in semi-major axis -- eccentricity ($a$--$e$) space with the pericenter curve (figures \ref{fig:clm effect} and \ref{fig:sim regolith capture combined a vs e}).

We illustrate the qualitative behavior of orbiting particles in our simulation in~\cref{fig:clm effect}. For a perfectly spherical central body, ejected particles will escape from or fall back onto the central body. Any particle orbiting the spherical central body will have the same orbit and show no change in $a$--$e$ space. However, the orbits are very quickly changed in the ellipsoidal Centaur's gravity field. Particles below the pericenter curve (region 1 in \cref{fig:clm effect}) associated with $R_{min}$ hover about their initial orbit with no major change. This makes region 1 a stable region. Particles far above the pericenter curve (region 3 in \cref{fig:clm effect}) eventually crash back onto the Centaur as their pericenter is below $R_{min}$. Particles along and close to the pericenter curve (region 2 of \cref{fig:clm effect}) are more chaotic as they ``shuttle" or ``bounce" along the curve. They tend to be excited and/or damped along the curve over the course of the simulation. However, they do not fall back onto the Centaur. The particles just above the curve but close to it are quickly ``pulled down" along the curve in the first $\sim 0.3$ rotation. Some of these particles fall underneath the curve and are caught in orbit. They then bounce around the curve but stay in orbit for the duration of our simulations.

The above analysis does not include the effect of inter-particle interactions. It shows that the non-spherical gravity field allows for temporary capture, but the inter-particle gravity and collisions stabilize the capture. Inter-particle gravity helps counteract the angular momentum increase from the azimuthal terms and slows particle escape. Collisions between regolith particles also damp the overall orbits. The collisional fragments are smaller and have damped orbits that places them under the pericenter curve. As a result, a large number of particles are de-excited and able to fall into the stable region under the pericenter curve, i.e., region 1 of \cref{fig:clm effect}.

\subsection{Gravitational Harmonics} \label{sec:shgrav results}

The above regolith capture is possible only because of the strong higher order gravitational harmonics terms ($C_{l,m}$) that alter the gravity field. For a perfectly spherical central body, we have a standard 2-body Keplerian gravity field and ejected particles will either escape or fall back onto the body. These terms arise from the ellipsoidal shape of the Centaur which implies azimuthal asymmetry. This means that using only the zonal terms ($m = 0$, such as $J_2$ and $J_4$) for modeling the ellipsoidal gravity field is not a good simplifying assumption. We have to include the azimuthal/tesseral terms ($m \neq 0$, such as $C_{2, 2}$) in our models. To explain this, we split the zonal and azimuthal terms, and test the effect of adding subsequent higher degrees. 

\begin{figure}[h]
    \centering
    \plottwo{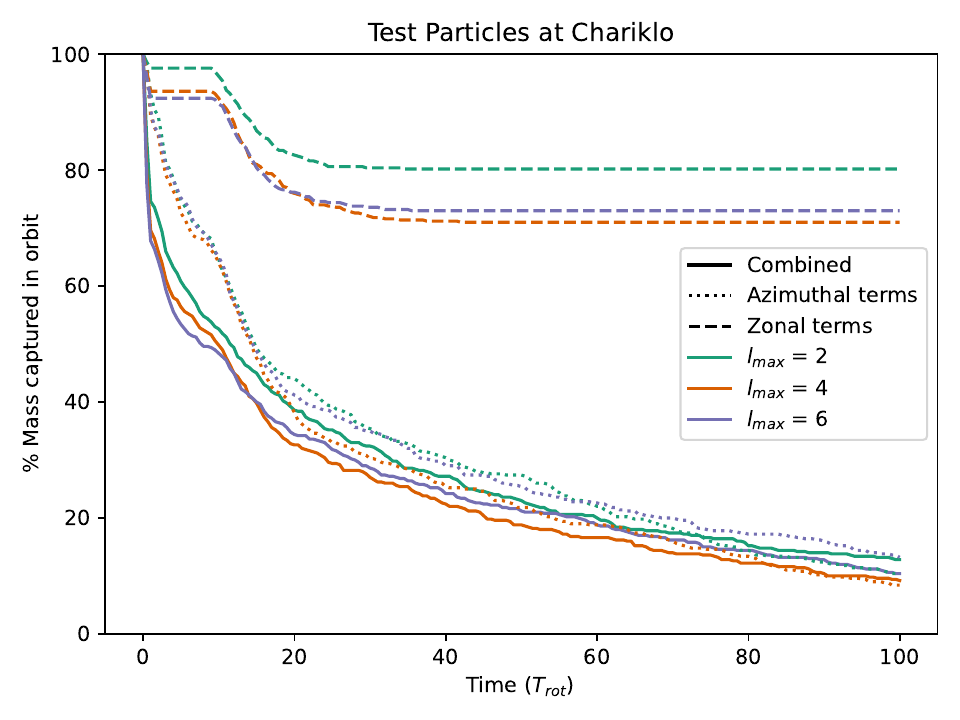}{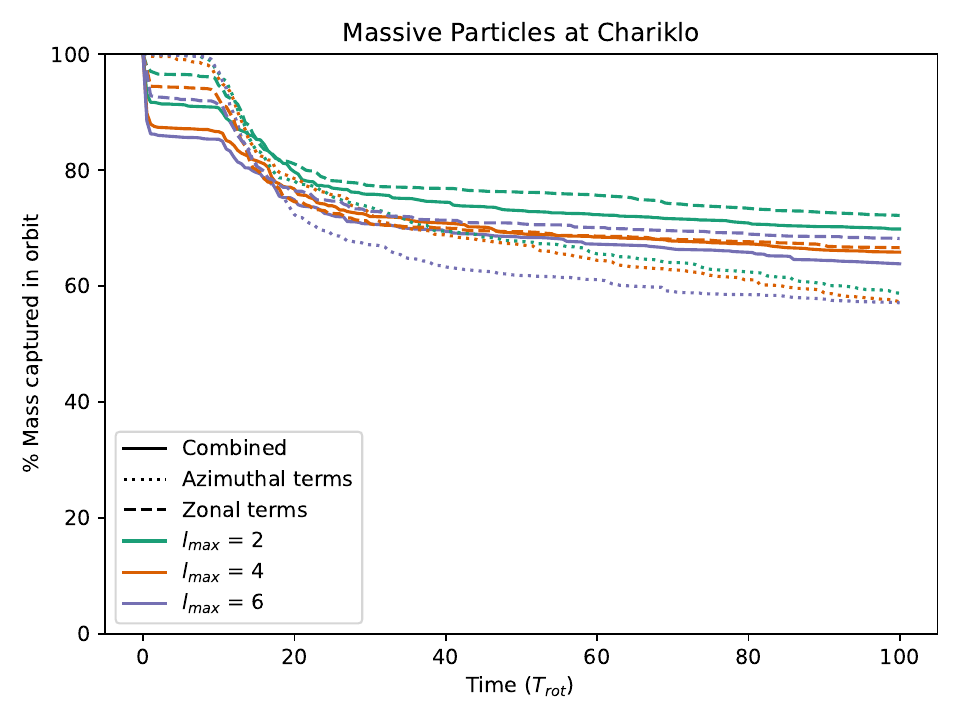}
    \caption{Regolith mass in orbit vs time when considering different gravitational harmonics terms for a landslide-like ejection around Chariklo. The simulations with only zonal terms (dashed lines) cause some initial mass loss but then do not drive regolith particles out of orbit. The azimuthal terms (dotted lines) dominate and drive particles out of the system because of angular momentum transfer between the Centaur and the regolith particles. The test particle simulations (left) clearly show this behavior, while the massive particle simulations (right) are more complex because of inter-particle interactions. Inter-particle gravity and collisions typically help counter the regolith particle loss.\label{fig:shgrav mass capture}}
\end{figure}

We run sets of simulations with only the zonal terms ($m = 0$), only the azimuthal terms ($m \neq 0$ and $|m| \leq l_{max}$ or $1 \leq |m| \leq l_{max}$), and all terms combined ($|m| \leq l_{max}$). Each set is done for a given maximum gravitational harmonic degree $l_{max}$. This means that each set has the $l = 0, 1, 2, ..., l_{max}$ terms included with $m$ bounded as above. We test this for $l_{max} = 2, 4, 6$. There is no asymmetry in the northern and southern hemispheres in the shape model used in~\citet{Leiva2017} and thus, the odd-degree/order terms are $0$. Regolith particles are ejected in a landslide-like ejection from the longest axis in the equatorial plane. In \cref{fig:shgrav mass capture}, we plot the regolith capture results for these simulations. 

\subsubsection{Azimuthal vs Zonal Terms} \label{sec:shgrav terms comparison}

From the test particle case (left-hand side of \cref{fig:shgrav mass capture}) we see that the azimuthal/tesseral terms ($1 \leq |m| \leq l_{max}$) and the zonal terms ($m = 0$) show vastly different behavior. The zonal terms look to stabilize the regolith in orbit and increase regolith capture. On the other hand, the azimuthal terms seem to drive them out of the system with angular momentum transfer. When all terms are combined, the behavior of the regolith particles largely follow that from the azimuthal terms. Therefore, the azimuthal terms dominate the zonal terms at all degrees ($l$).

However, for the massive particle case (right-hand side of \cref{fig:shgrav mass capture}) everything is more intertwined. The zonal terms show qualitatively similar behavior for all degrees with slightly higher regolith loss compared to the test particle case. This loss may be from collisional fragmentation. Inter-regolith interactions counteract the effect of the azimuthal terms, largely reducing regolith loss. This means that there is no clear domination of either azimuthal or zonal terms leading to a more comparable push and pull between their effects. This leads to higher regolith capture.

\subsubsection{Behavior Across Gravitational Harmonic Degrees} \label{sec:shgrav degree comparison}

Looking at the effect of each degree ($l$), there is most capture with $l_{max} = 2$, then $l_{max} = 6$, and then $l_{max} = 4$ for the test particle simulations across all types of harmonics terms. Compared to the $l = 2$ zonal term ($m = 0$; $C_{2, 0} = -J_2$), $C_{4, 0}$ captures $\sim 9\%$ lesser regolith, but $C_{6,0}$ captures $\sim 7\%$ lesser regolith. The $C_{6,0}$ term counteracts the $C_{4,0}$ term and captures an extra $\sim 2\%$ of regolith initially ejected. The azimuthal terms look to drive particles away but show similar behavior across degrees as above. The value of the gravitational harmonics coefficients ($C_{l, m}$) show the reason for this alternating behavior between subsequent degrees. $C_{2, 0}$ ($-J_2$) and $C_{6, 0}$ ($-J_6$) terms are negative  while the $C_{4, 0}$ ($-J_4$) is positive. This extends to higher $m$ values as well with the $l = 4$ coefficients having flipped signs compared to the corresponding values at $l = 2, 6$.

However, in the massive particle simulations, we have the most regolith capture for $l_{max} = 2$, then $l_{max} = 4$, and then $l_{max} = 6$. The difference in the latter two cases is $\sim 1.7\%$ lesser regolith in the combined-terms case. The reason for this difference is that the $l = 6$ azimuthal terms drive particles away faster than the $l=4$ terms in the initial stages. While the final capture amount in the azimuthal case is effectively the same for the two, it leads to slightly lesser capture in the combined-terms case. As alluded to earlier, the lack of domination of either type of term means that small changes can cause larger effects, leading to more varying effects. This clouds the effect of individual terms and while the $l = 6$ zonal term definitely helps with regolith capture, the $l = 6$ azimuthal terms may not.

From these results, we can infer that the higher degree and order gravitational harmonics have a clear effect on regolith capture and must be included for an accurate analysis for ellipsoidal bodies. Especially for test particle simulations, just using $J_2$ is not a good simplifying assumption.

\section{Discussion} \label{sec:discussion}

The above results show that ejected surface regolith can be captured in orbit around a Centaur as a disk for at least $100$ Centaur rotations. An initial landslide-like ejection from the equatorial plane with an ejection velocity of $0.2 - 0.5$~v$_{\mathrm{escape}}$ has the highest probability to place material in orbit. The gravity field of the ellipsoidal Centaur and regolith-regolith interactions alter and damp the regolith orbits to facilitate temporary capture. This regolith disk has a large eccentricity spread and may serve as proto-ring material. 

The biggest assumption of our work is the large regolith particles ($200 - 500$~m) to maintain computational efficiency. Test particle simulations of regolith are not particularly analogous to small regolith particles because we miss out on collisional fragmentation effects and inter-regolith gravitational effects. Collisions are crucial to place regolith particles in orbit. The gravitational interactions for small regolith are likely dominated by the central body, but would potentially counter the effect of the azimuthal gravitational harmonics terms to some extent (see \cref{fig:shgrav mass capture}). If the regolith particles were smaller and more realistic, we expect the same large-scale result of temporary capture. We would expect more collisions because of the increase in number density, lesser fragmentation, and potentially more collisional damping for the same total mass. However, we cannot comment on specifics about regolith size because the collisional fragmentation models in Swiftest are not constrained for bodies well into the strength regime~\citep[$R \lesssim 100$~m,][]{Stewart2009Velocity-DependentPlanetesimals}.

We eject regolith particles from the equatorial (x -- y) plane because they receive the highest centrifugal velocity kick ($\mathbf{v}_{rot} \simeq 0.45$~$\mathbf{v}_{escape}$ for Chariklo) in this plane. The position vector of a regolith particle ($\mathbf{r}$) is the longest in this plane and with $\bm{\omega}$ being in the $\mathbf{\hat{z}}$ direction, the cross product $\mathbf{v}_{rot}$ lies in the equatorial plane tangential to $\mathbf{r}$. Ejections from higher latitudes have a significantly reduced centrifugal kick as $\mathbf{r}$ decreases quickly with the ellipsoidal shape (longest axis goes from $157$~km to $86$~km for Chariklo; see table \ref{tab:centaurs}) and due to a smaller projection onto the x -- y plane from the cross-product. Radial outbursts are not conducive to orbit capture and thus, a significantly larger and faster landslide-like outburst is required. In addition to requiring a more specific direction, this also pushes the necessary ejection velocity values to the rarer observed values, and reduces the probability of a temporarily captured disk.

In this work we do not model the long-term behavior of the captured disk as it potentially evolves into the observed rings. Because of N-body computational constraints, the large regolith particles have low spatial density. As a result, intra-disk particle behavior, such as viscous spreading due to inter-particle collisions and the effect of subsequent ejections on an existing disk, would be modeled incorrectly. This is one of the reasons why we restrict our simulations to $100$~rotations ($\sim 29$~days for Chariklo and $\sim 25$~days for Chiron). We settled on this time limit to allow the regolith disk to reach a steady state but not enter deep into the ring evolution regime. The mass loss rate is relatively small by $100$~rotations. We see steady behavior from the captured particles after $\sim 15$~rotations, albeit with a slow mass loss. Simulations with higher capture efficiency tend to have a lower loss rate by the end of our simulations than simulations with lower efficiency. Therefore, our results may conservatively underestimate the difference between different cases.

Recent lab studies of collisions between icy ring-particles can give us viscosity estimates for small bodies and an approximate spreading timescales of $\sim 30$~yr for Chariklo and $\sim 60$~yr for Chiron~\citep{Brisset2024RingViscosity}. This is much longer than our simulations and helps us steer clear of incorrect results. While we do not model future behavior, we expect the disk to circularize and viscously spread over time depending on the collisional behavior of the icy ring particles~\citep{Brisset2024RingViscosity}. \citet{Winter2023OnRings} show that the $J_2 (-C_{2, 0})$ and $C_{2, 2}$ terms restrict circularized proto-ring material into a tight band and a ring-like structure, and our test simulations largely agree. This would be a potential avenue to counter the slow mass loss we see in \cref{fig:chariklo landslide mass capture combined,fig:chiron landslide mass capture}. In addition, the effect of subsequent ejections on the regolith disk is unknown and may vary from disrupting the disk or feeding material into it. However, further testing is required and the above is out of the scope of our project. 

Because only a limited number of active Centaurs have been seen, a ring or disk system could indicate past activity. For example, while no activity has been observed on Chariklo, a past landslide or avalanche-like outburst (and any subsequent outbursts) may have led to the rings currently observed. The ring system is likely advanced because of the lack of a coma, and this ejection had to have happened a long time ago. This implies that there are potentially more unobserved rings or regolith disks around currently dormant Centaurs.



\subsection{The Gap in Chariklo's Rings} \label{sec:Chariklo ring gap}

On the gap in Chariklo's rings, we offer two potential ways to combine current knowledge with our findings. The typical way to produce the gap is via shepherding satellites~\citep{Braga-ribas-2014, Sickafoose2024NumericalPerturber}. There is a varied size distribution of captured particles in the proto-ring disk. As the disk spreads, circularizes, and evolves, regolith particles may coalesce and form tiny moonlets because of the low critical Roche density of $\rho = 0.85$~g cm$^{-3}$~\citep{Tiscareno2013CompositionsDensity} in the region. The disk may follow the more conventional mechanism of moonlets forming at the outer edge of the disk, or potentially in the middle of the disk and sweep out a gap. Interactions with spin-orbit or other resonances, and/or the ellipsoidal gravity field may also have adverse effects for the ring-moon process. These moonlets could then act as shepherding satellites and create the gap we observe.

Another potential method of recreating this gap is by having an eccentricity gradient~\citep{Hahn2023NbodyRinglets, Murray2000SolarDynamics}. The captured regolith disk in our model has a large eccentricity spread. Potentially, the proto-ring disk does not circularize completely or as quickly, and the eccentricity gradient in the narrow disk maintains a gap in the rings.

\section{Conclusions} \label{sec:conclusions}

The origin of Centaur rings is currently unknown. In this work we show that ejected surface regolith can be a source material for Centaur rings. Because of the lack of probable external causes \citep{Melita2017}, we tested innate characteristics of Centaurs: cometary activity \citep{Bauer2008, Jewitt2024TheContinuum}, highly ellipsoidal shape, and fast rotation \citep{Fornasier2014TheActivity, Leiva2017}. A single landslide-like outburst with an initial velocity of $0.2 - 0.5$~v$_{escape}$ shows the highest probability of having material in orbit around the Centaur after $100$~rotations.
    
Centaur activity ejects surface regolith particles from the equatorial plane and the ellipsoidal shape creates an atypical gravity field with significant influence from the higher order gravitational harmonic terms. Together with the fast rotation and inter-regolith interactions, the ejected regolith particles are trapped in orbit for at least $100$~rotations. Inter-regolith gravity and collisions help damp their initial excited orbits and counter the angular momentum transfer from the tesseral gravitational harmonics terms to facilitate orbit capture. An initial avalanche or landslide-like ejection in the equatorial plane shows the highest probability of temporary capture into orbit. The resultant disk is maintained around the Centaur for at least $100$~rotations and potentially serves as a proto-ring disk. The dynamical evolution of this disk and its behavior with subsequent outbursts are not studied in this work, but are important to understand in the context of Centaur rings.

Not all observed Centaurs have noted activity, but the presence of a ring or proto-ring disk may indicate past activity. Chariklo's ring system is likely a late stage ring system with no coma but distinct rings, and Chiron's system is too complicated to draw clear inferences. We've shown that regolith ejections from a Chariklo-like and Chiron-like body get trapped in orbit around them, at least temporarily. A lot is still unknown about Centaurs such as their activity, the driving mechanisms, and longer-term ring dynamics. We hope this work helps answer some of the many questions about Centaur rings (especially Chariklo (10199)), their origins, and gives insights to general Centaur dynamics and their history.

\begin{acknowledgments} \label{sec:acknowledgements}

This work was supported by the NASA Solar System Workings Program, Grant 80NSSC20K0857. We would like to thank Dr. Kristel Izquierdo, Dr. Adrien Broquet, Dr. Matija \'Cuk, and Dr. Michael M. Sori for helpful conversations about the gravitational harmonics sections. 
We would also like to thank the two anonymous referees for greatly improving the manuscript. This work was partially supported by the NASA Solar System Workings Program grant \#80NSSC20K0857.

\software{Swiftest~\citep{Wishard2023Swiftest:Systems}, SHTools~\citep{Wieczorek2018SHTools:Harmonics}, SciPy~\citep{Virtanen2020SciPyPython}, NumPy~\citep{Harris2020ArrayNumPy}}
\facilities{Purdue Rosen Center for Advanced Computing (RCAC) Clusters: Negishi and Bell}

\end{acknowledgments}


\bibliography{references}{}
\bibliographystyle{aasjournal}



\end{document}